\documentclass[reprint,twocolumn,prx,floatfix,lengthcheck,citeautoscript,superscriptaddress,longbibliography]{revtex4-1}

\usepackage{amssymb,amsfonts,amsmath,amsthm}
\usepackage{graphicx}
\usepackage[english]{babel}
\usepackage[latin1]{inputenc}
\usepackage[pdftex,colorlinks=true,pdfstartview=FitH,linkcolor=blue,citecolor=blue,urlcolor=blue]{hyperref}

\usepackage{bm}
\usepackage{float}
\usepackage{mathtools}
\usepackage{color,soul}
\usepackage{times}
\usepackage{upgreek}
\usepackage{MnSymbol}
\usepackage{verbatim}
\usepackage[bottom]{footmisc}
\usepackage{bbold}

\usepackage[dvipsnames,svgnames,table]{xcolor}
\usepackage{hyperref}
\usepackage{fancyhdr}
\usepackage[english]{babel}
\usepackage[caption=false]{subfig}

\hypersetup{
pdfstartview={FitH},
colorlinks=true,    
linkcolor=NavyBlue, 
citecolor=Blue,   
filecolor=NavyBlue, 
urlcolor=NavyBlue   
}

\newcommand{\bea}{\begin{eqnarray}}
\newcommand{\eea}{\end{eqnarray}}
\newcommand{\be}{\begin{equation}}
\newcommand{\ee}{\end{equation}}

\newcommand{\ci}{\mathrm{i}}

\begin{document}
\title{Optomechanical parametric oscillation of a quantum light-fluid lattice}
\author{A.~A. Reynoso}
\affiliation{Centro At{\'{o}}mico Bariloche and Instituto Balseiro,
Comisi\'on Nacional de Energ\'{\i}a At\'omica (CNEA)- Universidad Nacional de Cuyo (UNCUYO), 8400 Bariloche, Argentina.}
\affiliation{Instituto de Nanociencia y Nanotecnolog\'{i}a (INN-Bariloche), Consejo Nacional de Investigaciones Cient\'{\i}ficas y T\'ecnicas (CONICET)-CNEA, Argentina.}
\author{G. Usaj}
\affiliation{Centro At{\'{o}}mico Bariloche and Instituto Balseiro,
Comisi\'on Nacional de Energ\'{\i}a At\'omica (CNEA)- Universidad Nacional de Cuyo (UNCUYO), 8400 Bariloche, Argentina.}
\affiliation{Instituto de Nanociencia y Nanotecnolog\'{i}a (INN-Bariloche), Consejo Nacional de Investigaciones Cient\'{\i}ficas y T\'ecnicas (CONICET)-CNEA, Argentina.}
\author{D. L. Chafatinos}
\affiliation{Centro At{\'{o}}mico Bariloche and Instituto Balseiro,
Comisi\'on Nacional de Energ\'{\i}a At\'omica (CNEA)- Universidad Nacional de Cuyo (UNCUYO), 8400 Bariloche, Argentina.}
\affiliation{Instituto de Nanociencia y Nanotecnolog\'{i}a (INN-Bariloche), Consejo Nacional de Investigaciones Cient\'{\i}ficas y T\'ecnicas (CONICET)-CNEA, Argentina.}
\author{F. Mangussi}
\affiliation{Centro At{\'{o}}mico Bariloche and Instituto Balseiro,
Comisi\'on Nacional de Energ\'{\i}a At\'omica (CNEA)- Universidad Nacional de Cuyo (UNCUYO), 8400 Bariloche, Argentina.}
\affiliation{Instituto de Nanociencia y Nanotecnolog\'{i}a (INN-Bariloche), Consejo Nacional de Investigaciones Cient\'{\i}ficas y T\'ecnicas (CONICET)-CNEA, Argentina.}
\author{A.~E. Bruchhausen}
\affiliation{Centro At{\'{o}}mico Bariloche and Instituto Balseiro,
Comisi\'on Nacional de Energ\'{\i}a At\'omica (CNEA)- Universidad Nacional de Cuyo (UNCUYO), 8400 Bariloche, Argentina.}
\affiliation{Instituto de Nanociencia y Nanotecnolog\'{i}a (INN-Bariloche), Consejo Nacional de Investigaciones Cient\'{\i}ficas y T\'ecnicas (CONICET)-CNEA, Argentina.}
\author{A.~S. Kuznetsov}
\affiliation{Paul-Drude-Institut f\"{u}r Festk\"{o}rperelektronik, Leibniz-Institut im Forschungsverbund Berlin e.V., Hausvogteiplatz 5-7,\\ 10117 Berlin, Germany.}
\author{K. Biermann}
\affiliation{Paul-Drude-Institut f\"{u}r Festk\"{o}rperelektronik, Leibniz-Institut im Forschungsverbund Berlin e.V., Hausvogteiplatz 5-7,\\ 10117 Berlin, Germany.}
\author{P.~V. Santos}
\affiliation{Paul-Drude-Institut f\"{u}r Festk\"{o}rperelektronik, Leibniz-Institut im Forschungsverbund Berlin e.V., Hausvogteiplatz 5-7,\\ 10117 Berlin, Germany.}
\author{A. Fainstein}
\email[Corresponding author, e-mail: ]{afains@cab.cnea.gov.ar}
\affiliation{Centro At{\'{o}}mico Bariloche and Instituto Balseiro,
Comisi\'on Nacional de Energ\'{\i}a At\'omica (CNEA)- Universidad Nacional de Cuyo (UNCUYO), 8400 Bariloche, Argentina.}
\affiliation{Instituto de Nanociencia y Nanotecnolog\'{i}a (INN-Bariloche), Consejo Nacional de Investigaciones Cient\'{\i}ficas y T\'ecnicas (CONICET)-CNEA, Argentina.}
\date{\today}

\begin{abstract}
Two-photon coherent states are one of the main building pillars of non-linear and quantum optics. It is the basis for the generation of minimum-uncertainty quantum states and entangled photon pairs, applications not obtainable from standard coherent states or one-photon lasers. Here we describe a fully-resonant optomechanical parametric amplifier involving a polariton condensate in a trap lattice quadratically coupled to mechanical modes. The quadratic coupling derives from non-resonant virtual transitions to extended discrete excited states induced by the optomechanical coupling. Non-resonant continuous wave ($cw$) laser excitation leads to striking experimental consequences, including the emergence of optomechanically induced  inter-site parametric oscillations and inter-site tunneling of polaritons at discrete inter-trap detunings corresponding to sums of energies of the two involved mechanical oscillations ($20$ and $60$ GHz confined vibrations). We show that the coherent mechanical oscillations correspond to parametric resonances with threshold condition different to that of standard linear optomechanical self-oscillation. The associated Arnold tongues display a complex scenario of states within the instability region. The observed new phenomena can have applications for the generation of entangled phonon pairs, squeezed mechanical states relevant in sensing and quantum computation, and for the bidirectional frequency conversion of signals in a technologically relevant range.
\end{abstract}
\maketitle
\section{Introduction}
The concept of two-photon coherent states associated with quadratic Hamiltonians has been a revolutionary development in quantum optics~\cite{Gerry-Knight2004}. They can be obtained from normal coherent states through unitary operators associated with quadratic Hamiltonians \cite{Yuen1976,Yuen-Shapiro1979}. They correspond to the radiation states of ideal two-photon lasers operating far above threshold.  In real devices, this is accomplished by using parametric processes in materials with large second order susceptibilities \cite{Slusher1985,Wu1986}.  To enhance such high-order processes, particularly in continuous wave operation, optical parametric oscillators (OPOs) exploit a resonator so that the laser pump and either one or both parametrically generated photons (signal and idler) are confined, leading to feedback in multi-wavelength cavities \cite{Yang1993,Zeng2014}. The emission of coherent entangled pairs of photons from such OPOs is at the base of different approaches for  autocorrelation ($g^{(2)}$) measurements, Bell test experiments, and quantum key distribution protocols in quantum communications ~\cite{Kimble2014}. The mean-square quantum noise behaviour of these states, which is basically the same as those of minimum-uncertainty states, leads to applications in sensing not obtainable from one-photon lasers \cite{Bondurant1984,Polzik1992,Lawrie2019}. Moreover, such quantum many-photon squeezed vacuum states are involved in the most recent and promising approaches to large scale photonic quantum computing, as an alternative to implementations using single photons \cite{Silberhorn2017,Zhong2020}.

Phonons, the quanta of mechanical vibrations, can also be used to define and manipulate coherent states. Compared to cavity photons, acoustic phonons have a longer lifetime and a much smaller wavelength (which scales with the relation of sound to light velocity), two features relevant for device integration.  These properties are exploited, for example, in microwave-to-sound conversion and filter components based on acoustic waves, as extensively used in cell phones~\cite{Ruppel2017} but also with interesting prospects in quantum technologies \cite{SAW-Train2019}.  Vacuum squeezing of the mechanical motion of solids has been reported by driving with femtosecond laser pulses materials characterised by large two-phonon Raman scattering susceptibilities \cite{Garret1997,Nori1999}.  More recently, the search for quadratic mechanical Hamiltonians (ie., displaying a coupling quadratic in the displacement, $x^2$) has gained renewed interest in the framework of the rapidly developing field of cavity optomechanics \cite{RMP}.  A number of cavity-optomechanical systems  have been explored \cite{Painter2015}, the membrane-in-the-middle approach being probably the most paradigmatic platform~\cite{Thompson2008} but also with other realizations including systems of cold atoms and levitated nanoparticles \cite{Purdy2010,Bullier2020}. The $x^2$ coupling has been proposed as a means for realising quantum nondemolition measurements of phonon number~\cite{Miao2009,Dellantonio2018},  measurement of phonon shot noise~\cite{Clerk2010}, and the cooling and squeezing of mechanical motion~\cite{Bhattacharya2008,Nunnenkamp2010,Wollman2015,Ma2021}.

Here, we describe a new concept for the generation of two-phonon coherent states based on an optomechanical crystal of exciton-polaritons (strongly coupled light-matter particles). Exciton-polaritons behave as a quantum light-fluid (LF)~\cite{CarusottoRMP2013}, showing Bose-Einstein condensation~\cite{Kasprzak2006} with peculiar features stemming from their composite boson nature, effective photon-photon interactions and the intrinsically driven-dissipative non-Hermitian character of their dynamics. Besides this, polariton condensates are interesting in the domain of cavity optomechanics due to their long coherence times, and
huge electrostrictive (exciton-mediated) optomechanical interactions \cite{Cerda-Mendez2010,Rozas2014,Jusserand2015,Kuznetsov2021}.

We present both experimental and theoretical results for an array of micrometer-size exciton polariton traps on which recently polariton-driven phonon lasing has been reported \cite{Chafatinos2020}. We show that upon non-resonant $cw$ excitation such a LF optomechanical lattice evolves into a stationary coexistence of condensed phases populating both the strongly localized ground state of neighbor traps, and a shared delocalized excited state. The energy detuning between neighbor traps depends on the exciton reservoir because of the involved Coulomb interactions. When this detuning becomes resonant with the combined energy of pairs of phonons confined in the resonators, an optomechanical instability arising from a quadratic coupling involving the excited state sets in. The observed instability corresponds to a parametric resonance with properties fundamentally different from self-oscillation typically observed in linear-coupling resonantly driven optomechanical systems. The physics is rather that of a parametric resonator and of two-phonon coherent states conceptually equivalent to two-photon resonant OPOs well above laser threshold. Concomitant with what we define as an optomechanical parametric oscillation state, a locking of the polariton trap energies, and an enhanced excited-state mediated optomechanically induced tunneling are observed.
\begin{figure*}[!hht]
 \begin{center}
    \includegraphics[trim = 20mm -10mm 0mm 0mm,scale=0.3,angle=0]{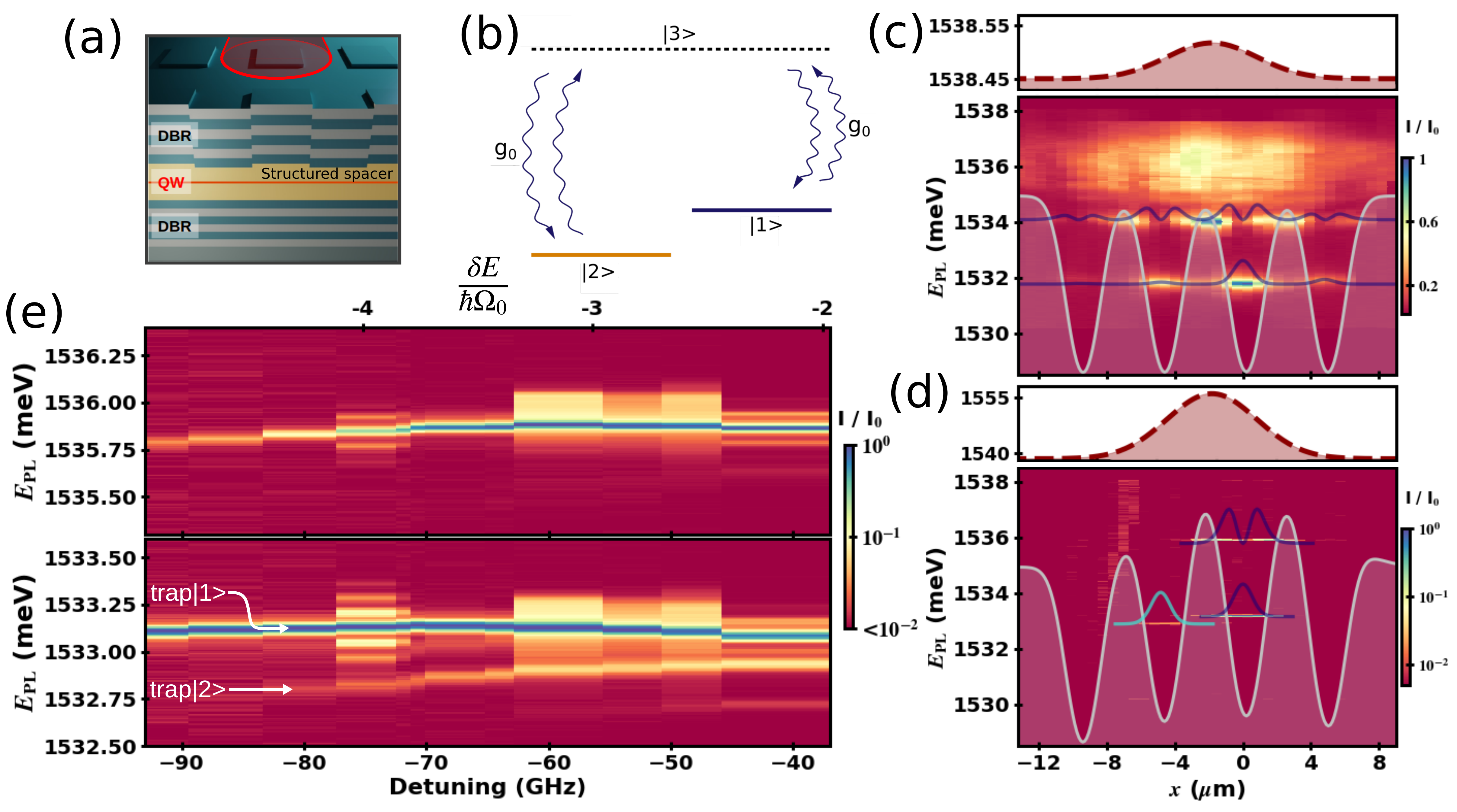}
\end{center}
\vspace{-0.8 cm}
\caption{
\textbf{Polariton effective potentials and optomechanical sidebands.}
Panel (a) is a scheme of the lattice consisting of a pumped central polariton trap and neighbor traps in a square  array. Panel (b) presents the involved polariton levels corresponding to the strongly confined ground states of the pumped trap ($\left| 1 \right>$), and one of its neighbors traps ($\left| 2 \right>$), as well as a  delocalized  excited state ($\left| 3 \right>$). $g_\mathrm{0}$ is the linear optomechanical interaction coupling the $s$-symmetric ground states with the $p$-symmetric excited state with the concomitant emission of a $p$-like confined phonon. Panel (c) [(d)]
displays the spectral and spatial image corresponding to the low [high]-pump power condition. The effective trap potentials (shaded light red), lateral distribution of the exciton energy (top dashed Gaussian curve), and corresponding confined polariton wavefunctions (thin coloured curves) derived from a Gross-Pitaevskii modelling of the quantum light fluid are also shown.
The detuning between the pumped and neighbor trap ground states ($\delta E$) can be tuned through the excitonic-related repulsive interaction with the reservoir by varying the non-resonant incoherent pump power as shown in panel (e) (note that this detuning is given in units of the phonon energy $\hbar\Omega_0$ in the top scale). The emission from ground states $\left| 1 \right>$, $\left| 2 \right>$, and the excited state $\left| 3 \right>$ can be identified.
Note that clear and symmetrical spectral sidebands appear for both the ground and excited state precisely when one of the neighbor traps is red-detuned with respect to the pumped trap by integer multiples $2$ and $4$ of the fundamental breathing mechanical cavity mode $\nu_\mathrm{m}^\mathrm{0} \sim 19$~GHz.
}
\label{Fig1}
\end{figure*}

\section{Results}
\paragraph*{\textbf{The light fluid lattice.}} The studied system consists of cavity polaritons in $\mu$m-sized intracavity traps patterned by means of a shallow etch-and-overgrowth technique on a (Al,Ga)As microcavity~\cite{Kuznetsov2018}. The lateral  modulation of the spacer of the microcavities creates an array of intracavity traps, each consisting of a non-etched area surrounded by etched barriers (a detailed description of the device is presented in Appendix~\ref{Fab}). These layered cavity structures also confine breathing-like vibrations polarized along the growth direction ($z$), with a fundamental frequency around $\nu_\mathrm{m}^\mathrm{0} \sim 20$~GHz and overtones at  $\nu_\mathrm{m}^n = (1+2n) \nu_\mathrm{m}^\mathrm{0}$, $n=0,1,2...$ (the first one being at $\nu_\mathrm{m}^\mathrm{1} = 3\nu_\mathrm{m}^\mathrm{0}  \sim 60$~GHz) \cite{Fainstein2013}. The lateral patterning adds an additional trapping potential for acoustic phonons in a way fully equivalent to the confinement of photons.
Thus, similar to polaritons, acoustic phonons can be described as confined modes of in-plane $s$, $p$, $d$.... -like symmetry~\cite{Anguiano2017}.  We will concentrate here on experiments performed on an array of square traps of $1.6~\mu$m lateral size separated by $3.2~\mu$m-wide etched barriers (a scheme is shown in Fig.~\ref{Fig1}(a)).

A typical spatially resolved photoluminescence (PL) image obtained below and above the polariton condensation threshold is presented in Figs.~\ref{Fig1}(c) and \ref{Fig1}(d), respectively.  The roughly $3~\mu$m-wide  Gaussian-like laser spot mainly excites a single trap. Neighboring traps are, however, also weakly excited through the tails of the laser spot as well as via lateral propagation of the excitons in the reservoir. A small unintentional misalignment of the laser spot on the addressed microstructure typically leads to one of the neighbor traps being more strongly excited than the others (the one in the left for the shown example). The fundamental and first excited states of the pumped trap, as well as weaker contributions from neighbor traps, can be identified in Figs.~\ref{Fig1}(c) and \ref{Fig1}(d). The transition to a LF condensed state is signaled both by the narrowing of the lines and the increase of the emitted intensity from the lower confined levels~\cite{Kuznetsov2018}. Above threshold narrow coherent emission can be identified mainly from the ground state, but also from the first excited state.

The spatial and pump power dependence of the strongly-coupled polariton states in the traps can be described by a phenomenological Gross-Pitaevskii equation (GP from now on). Since all the energy scales involved are of the same order of magnitude  (exciton-photon detuning, Rabi splitting, and correlations-induced blue-shift), our GP model allows for a change in the exciton-photon content as a function of the excitation power  and includes the spatial modulation of the photon cavity mode energy introduced by the microstructuring of the cavity spacer (see  Appendix \ref{GP}) \cite{Mangussi2021}. From this model we can obtain the effective potential affecting the polaritons in the traps, as well as the energies and spatial distribution of the trap states.
Figures~\ref{Fig1}(c) and \ref{Fig1}(d) show, together with the experimental data, the calculated polariton effective potential, confined trap energies and polariton modes. In the latter a significant detuning between neighbor trap states  develops due to the spatially dependent blue-shift induced mainly by the interaction with the reservoir. The agreement with the experiments is notably good. One important information derived from these effective potentials are the inter-trap overlaps for the different states and applied powers. For the overlap integral between ground states of neighbor traps we obtain a very small value $ \sim 10^{-4}$, highlighting a very strong localization of the ground polariton state in individual traps. This contrasts with the overlap integral between excited states, for which we obtain $ \sim 0.4$, evidencing that both at low and high powers the first excited state corresponds to a mode spread out in both traps.

\begin{figure*}[!hht]
    \begin{center}
    \includegraphics[trim = 0mm 0mm 0mm 0mm, clip=true, keepaspectratio=true, width=1.5\columnwidth, angle=0]{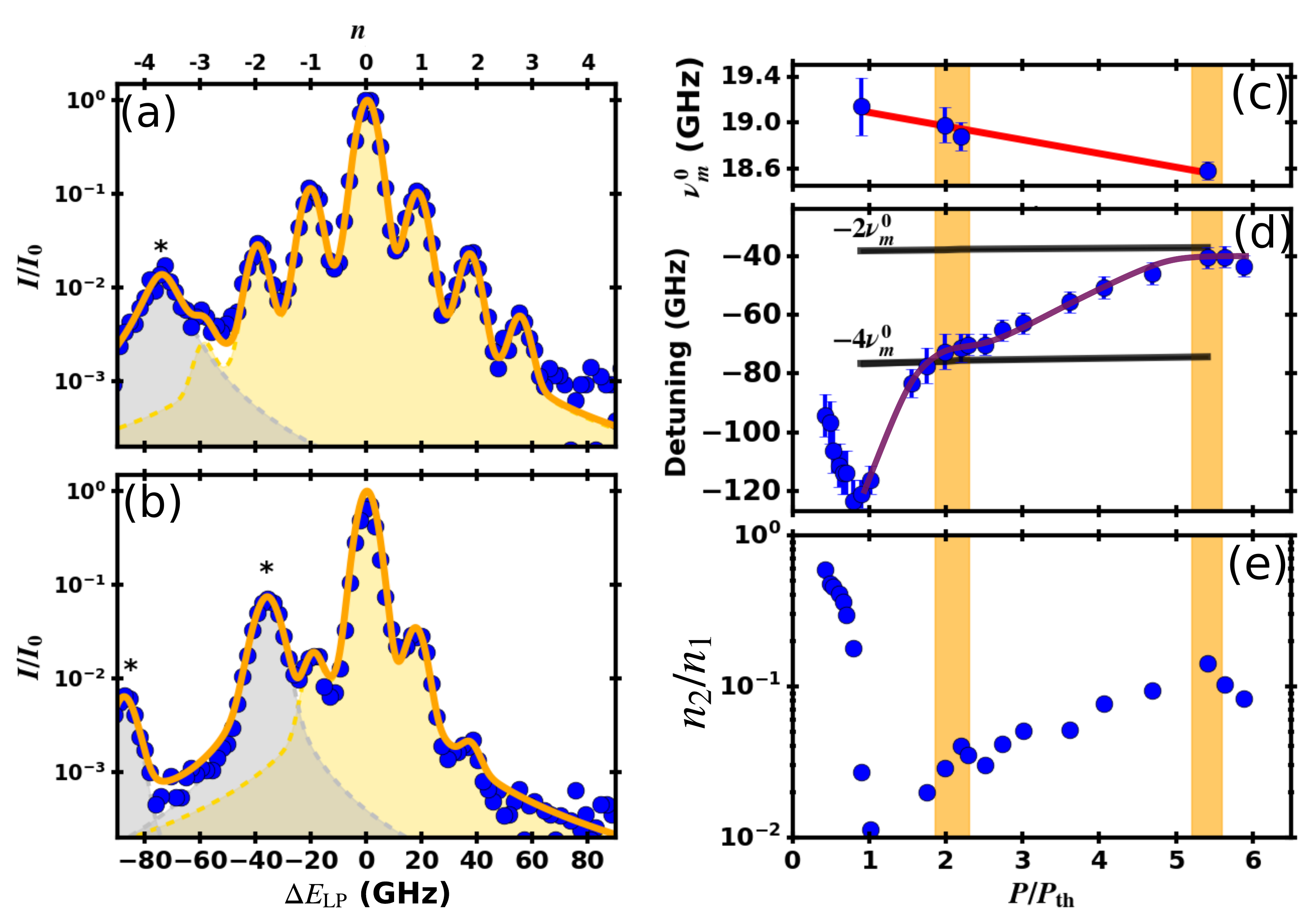}
    \end{center}
\caption{{\bf Mechanical mode softening, inter-trap frequency locking, and enhanced polariton transfer.}
Examples of PL emission precisely at the resonant detuning conditions $\delta E_{4} \sim 75$~GHz and $\delta E_{2} \sim 37$~GHz are shown in panels (a) and (b), respectively. Asterisks indicate PL from neighbor traps. The symmetric evenly spaced mechanical sidebands can be clearly observed at detunings corresponding to integer numbers $n=2$ and $n=4$ of the fundamental mechanical frequency $\hbar\Omega_0$. Fits to the spectra (indicated with continuous orange curves in (a) and (b) allow for the derivation of the
applied incoherent non-resonant $cw$ pump power dependence of the involved mechanical frequencies, the trap energy-detuning, and their occupation (shown in panels (c)-(e), respectively). The vertical yellow bands indicate the regions where the inter-trap resonances are attained, leading to the observation of the mechanical sidebands.
}
\label{Fig2}
\end{figure*}

\paragraph*{\textbf{The optomechanical sidebands.}} The arrays studied here display, with increasing non-resonant \textit{cw} excitation power, rich optomechanical phenomena including strong phonon assisted PL and, upon stronger excitation, very efficient mechanical coherent oscillation (phonon lasing) \cite{Chafatinos2020}. The latter is evidenced by the appearance of well-resolved symmetrical sidebands, separated by the energy of the fundamental mechanical breathing mode of the resonators, for both the ground and excited state  emission (shown in Fig.~\ref{Fig1}(e)). Quite notably, the mentioned mechanical sidebands, signalling the passage to a coherent phonon state, appear when one of the neighbor traps is red-detuned with respect to the pumped trap by an integer multiple of the fundamental breathing mechanical cavity mode energy $h\nu_\mathrm{m}^\mathrm{0}$ \cite{Chafatinos2020}. More precisely, they occur at  $\delta E_{4} = -4h\nu_\mathrm{m}^\mathrm{0}$ ($\sim 75$~GHz) and $\delta E_{2} = -2h\nu_\mathrm{m}^\mathrm{0}$ ($\sim 37$~GHz)---cross-sections for these detunings are shown in Figs. \ref{Fig2}(a) and \ref{Fig2}(b), respectively. These observations clearly point to a resonant two-mode optomechanical system, involving the polariton ground states of the pumped and a neighbor trap, and the mechanical breathing oscillation of the optical cavities. We note that a strong modification of both the ground and excited state spectra in Fig.~\ref{Fig1}(e) can also be observed for detunings $\delta E \sim 50-60$~GHz ($\sim3\hbar\Omega_0$). The states in this case, however, do not display symmetrical well-resolved sidebands but an asymmetric distortion of a different nature.

The standard approach for optomechanical linear coupling (cavity-phonon coupling proportional to $x$)
based on a two-mode cavity system leads to the optomechanically modified phonon effective lifetime $\Gamma_\mathrm{eff}=\Gamma_\mathrm{m} (1-C)$, where $C$ is the so-called optomechanical cooperativity \cite{RMP}. This result holds when the two polariton modes of energy $\hbar \omega_i$ are detuned by the phonon energy $\hbar \Omega_0\equiv h \nu_\mathrm{m}^0$, that is, for  $\omega_1-\omega_2=\Omega_0$. We use here the trap index $\{1,2 \}$, where $1$($2$) refers to the pumped (neighbor) trap.  To derive $\Gamma_\mathrm{eff}$  a strong coherent driving tuned to the blue-shifted polariton mode $1$ is also assumed while cavity losses are included through the linewidths $\kappa$ of the polariton traps, and $\Gamma_\mathrm{m}$ of the involved mechanical vibration. It follows that for such two-mode resonant linear optomechanical systems, the threshold for self-oscillation is reached provided that $1<C=4 \frac{N_{\mathrm{1}} |g|^2}{\kappa\Gamma_\mathrm{m}}$, where $g$ is the optomechanical single polariton coupling rate and $N_1$ the occupation of the driven mode.

The optomechanical factor coupling polariton levels of the same trap can be estimated to be in the range $g/2\pi \sim 0.05-5$~MHz \cite{Fainstein2013,Kuznetsov2021,Villafane2018}, depending on the degree of radiation pressure or resonant electrostrictive contribution to the coupling
(see Appendix \ref{gfactor})
---from hereon we refer to this value as $g_0$ to emphasise its on-site character. For the two-mode situation described by our experiments, however, the optomechanical coupling $g$ connecting ground states of separate traps should be much smaller because of the mentioned overlap integrals being of the order of $10^{-4}$.
Confined LFs act as an intra-cavity coherent source.
It turns out, however, that because of the mentioned strongly isolated character of the polariton ground states, and being  $C \propto \left| g\right|^2$, the optomechanical cooperativity based on these calculations is several orders of magnitude smaller that the one required to account for the observed threshold to mechanical self-oscillation, even considering the upper limit of resonant electrostrictive contribution. Moreover, we note that the phonon sidebands are observed when the detuning between traps is \textit{not} $\hbar \Omega_0$, but am even multiple ($n=2, 4$) of it.

\paragraph*{\textbf{Mechanical mode softening, inter-trap frequency locking, and enhanced tunneling.}}
We now address additional features that arise in connection with the resonant inter-cavity detunings $\delta E_4$ and $\delta E_2$. Figures \ref{Fig2}(a) and \ref{Fig2}(b) present the emission spectra corresponding to these two resonance conditions in which symmetric evenly spaced mechanical sidebands are observed. The asterisks in these panels indicate peaks related to PL coming from neighbor traps (also highlighted with a darker shading).  Phenomenological lorentzian fits to the main lines and phonon-related sidebands in the polariton emission spectra allow for the determination of the phonon frequency, the inter-trap detuning, and the relative ground state occupation of the neighbor traps when the power of the non-resonant pump is varied.
These three features are displayed in panels (c)-(e) of Fig.~\ref{Fig2}. Vertical orange bands identify the powers at which mechanical sidebands are observed. There is a softening of the phonon frequency as the power is increased (panel (c)), amounting to about $1-2 \%$ over the whole scanned power range. Such softening resembles the so-called optical spring effect typical of back-action in cavity optomechanical phenomena \cite{RMP}. The inter-trap detuning shown in panel (d) is in principle determined by the distinct blue-shift of the pumped and the neighbor trap ground states induced by interactions between confined LFs and the excitonic reservoir. Quite notably, close to the first mechanical instability regime at $\delta E_4 \sim 75$~GHz, the rapidly evolving detuning changes its slope, partially locking around this value for a range of applied powers. After a further increase of the applied power, the reappearance of a mechanical coherent state coincides with the final locking of the detuning at  $\delta E_2\sim 37$~GHz. This behavior, indicative of a locking of the inter-trap energy detuning at an energy scale defined by the optomechanics, is highlighted in Fig.~\ref{Fig2}(d) with a guide-to-the-eye continuous curve superimposed on the data.

The information concerning the inter-trap polaritons transfer, on the other hand, is reflected in the intensity of the lines in the emission spectra as a function of increasing power (i.e., as a function of trap detuning), as displayed in Fig.~\ref{Fig2}(e), where the relative intensity $n_2/n_1$ is shown (note the log scale). Note that maxima of $n_2/n_1$ are observed when the instability to a mechanical coherent state occurs, bringing evidence of an enhanced polariton tunneling from the pumped to the neighbor trap when the peculiar resonant conditions are met.
The origin of such resonant mechanical instability at detunings corresponding to multiples of the phonon frequency, the mechanical mode softening, the inter-trap locking, and the optomechanically induced polariton tunneling will be addressed next.
\paragraph*{\textbf{Model Hamiltonian for a trap array with optomechanical coupling mediated by an excited state.}}
To describe the appearance of optomechanical induced sidebands on the PL spectrum of the non-resonantly pumped trap-array, we introduce here a simplified model that captures the main physical ingredients. The model takes into account the two fundamental polaritonic modes of two neighboring cavities, a single  polariton excited state shared by both traps and an on-site phonon-mediated coupling between ground and excited states.
The Hamiltonian then has two contributions, $H=H_0+H_\mathrm{OM}$. Here
\be
H_0=\sum_{j=1}^3 \hbar\omega_j\,\hat{a}_j^\dagger\hat{a}_j^{}+\sum_{n} \hbar\Omega_{ n}\,\hat{b}_{n}^\dagger\hat{b}_{ n}^{ }\,,
\ee
describes the decoupled polariton and phonon modes: i)   $\hat{a}_j^{\dagger}$ ($\hat{a}_j$) creates (annihilates) a polariton in the $j$-mode with energy $\hbar\omega_j$, where  $j=3$ refers to the excited mode; ii) $\hat{b}_{n}^{\dagger}$  ($\hat{b}_{n}$) creates (annihilates) a $p$-phonon in the ${n}$-mode with energy $\hbar\Omega_{n}$.  The index $n$ labels the fundamental and the overtone mechanical modes so that, for example,  $\Omega_1=3\Omega_0 \sim 2\pi \times 60$GHz (for simplicity, we take $\Omega_0= 2\pi \times 20$GHz). The linear optomechanical coupling reads
\bea
H_\mathrm{OM}&=&\sum_{j=1}^2\sum_n \hbar g_{jn}(\hat{a}_j^\dagger \hat{a}_3^{}+\hat{a}_3^\dagger \hat{a}_j^{})(\hat{b}_{n}^\dagger+\hat{b}_{n}^{})
\,.
\eea
Note that there is no direct coupling between $\hat{a}_1$ and $\hat{a}_2$ but only with the excited mode. This is so because tunneling between the fundamental modes is negligible, and so is the direct optomechanical coupling constant. Because of the fully symmetric $s$ character of the traps ground state, and the $p$ symmetry of the polariton excited state, only acoustic phonons of $p$ symmetry need to be considered---$s$ modes couple to the number operator $\hat{a}_j^\dagger\hat{a}_j^{}$ and their role will be discussed below (see also the discussion in Appendix \ref{full_model}). In addition, we note that while the $s$-like mechanical vibrations turn out to be fully confined,  the $p$ phonon modes are extended and so shared between traps (as the excited polariton state).

It is straightforward to derive from the above Hamiltonian the equations of motion for $\hat{a}_j$ and for the dimensionless phonon displacement operator, $\hat{x}_n=\hat{b}_{n}^\dagger+\hat{b}_{n}^{}$.
Here we restrict ourselves to the semi-classical approximation, where the bosonic operators are replaced by complex functions, which implies that the solution of interest contains a large number of both polaritons and phonons and so quantum fluctuation can be ignored.
The dynamics of the system is then given by the set of coupled non-linear equations detailed in Appendix \ref{full_model}.
In order to describe the effective non-Hermitian dynamics we added a phenomenological term in the equations to account for the incoherent driving of the polariton modes induced by the excitons reservoir (see Eq. \eqref{Eq:full_model}).
Such a stimulated driving leads to an effective decay rate for the polaritons $\tilde{\kappa}_{j}=\kappa_{j}[1-(P_j/P_{\mathrm{th},j})/(1 + |a_j|^2\bar{R}/{\gamma_R} )]$, that describes the condensation of each mode when the rate $P_j$ (proportional to the creation rate of excitons in the reservoir, and hence to the laser power) is larger than the threshold $P_{\mathrm{th},j} \equiv \kappa_j \gamma_R/\bar{R}$. Here, $\bar{R}$ is the rate of stimulated scattering from the reservoir to the polariton modes. In the absence of optomechanical effects each mode would condense ($\tilde{\kappa}_{j}=0$) to an occupation $n^0_j = (P_{j}/P_{\mathrm{th},j}-1)n_0$ with $n_0=\gamma_R/\bar{R}$.  Additionally, we also include in the equations for $x_n$ a dissipative term proportional to $\Gamma_{n}$ (frequency linewidth) to account for the decay of the phonon modes.
The derived non-linear coupled equations were solved numerically for different values of the parameters using the Runge-Kutta method choosing initial conditions with low phonon number. In this way we obtain steady state solutions that are accessible when the system evolves from the low phonon occupied undriven condition. Attractors and multistabilities arising at \emph{higher} number of phonons~\cite{Marquardt2006} are not within the scope of the present work. We used the following parameters: $g_{j1}=3\,g_{j2}=2\times10^{-4}$, $\kappa_j=0.2$, $\Gamma=5\times10^{-4}$, in units of $\Omega_0$, and  $n_0=10^6$.
\begin{figure*}[!hht]
 \begin{center}
      \includegraphics[width=1.05\textwidth]{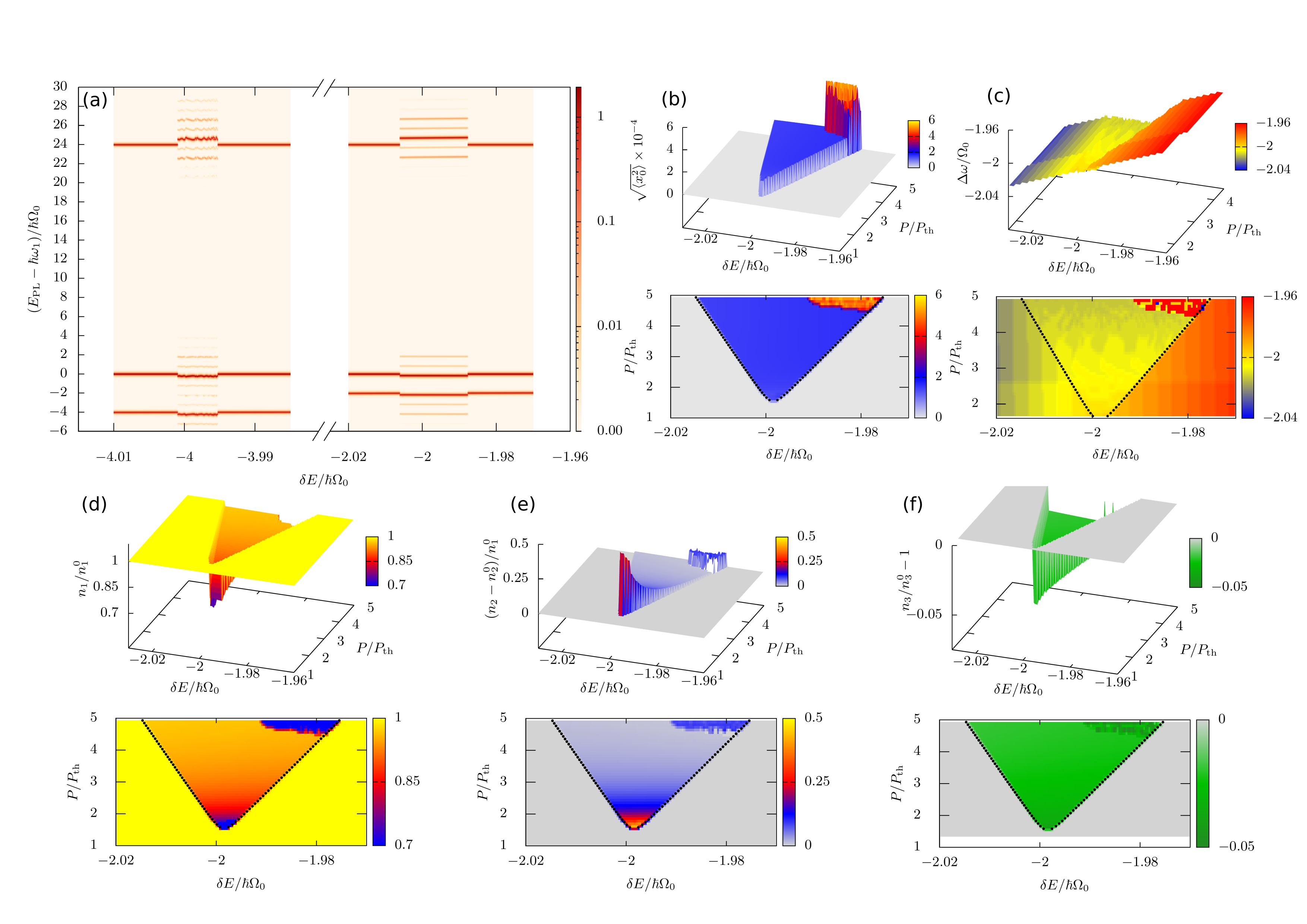}
\end{center}
\vspace{-0.6 cm}
\caption{\textbf{Optomechanical instability and Arnold tongues of the effective parametric resonator.}
Panel (a) shows the calculated PL spectra for the Hamiltonian $H_0+H_\mathrm{OM}$, as a function of the bare detuning $\delta E$ between traps $1$ and $2$. Here we set: $\omega_1=0$, $\omega_2=\delta E/\hbar$, $\omega_3=24\Omega_0$, and the threshold power as indicated in the main text.
Optomechanical instabilities are observed at $\delta E= -4\hbar\Omega_0$ and $\delta  E= -2\hbar\Omega_0$, corresponding to resonances involving the fundamental and first overtone of the mechanical vibrations ($\Omega_1\pm\Omega_0$), or two fundamental mechanical vibrations ($2\Omega_0$), respectively. The resolved additional peaks appear around both the ground and excited states, and are separated by the frequency of the fundamental mode ($\Omega_0$).
As a function of trap-detuning, and depending on the occupancy of the involved levels $n_\mathrm{1}$ and $n_\mathrm{2}$, instability (stability) regions can be defined characterised by the presence (absence) of coherent mechanical oscillations. The instability regions correspond to so-called Arnold tongues of a parametric resonator, with characteristic stationary values of phonon number [(b)], the dressed cavity detuning $\Delta \omega$ [(c)] and the traps level occupation, as displayed in panels (d)-(f). Note in (d) and (e) the strong optomechanically enhanced transfer of polaritons from trap $1$ to $2$, and in (c) the locking of their relative detuning within the instability region to the resonant value given by the sum of the two involved vibrations. Dotted lines in the colormap indicates the boundary of the Arnold tongue as estimated from Eq.  \eqref{threshold_condition}.
}
\label{Fig3}
\end{figure*}

Figure~\ref{Fig3} summarizes the most important results derived from this minimal model. Panel (a) presents the polariton spectrum above the threshold for optomechanical instability (to be discussed below) as a function of the detuning $\delta E/\hbar=\omega_2-\omega_1$ between the pumped and neighbor traps. Quite notably, sidebands separated by the fundamental mechanical frequency $\Omega_0$, occur at  $\delta E_4 \sim -4\hbar\Omega_0$ and $\delta E_2 \sim -2\hbar\Omega_0$, signalling the emergence of mechanical oscillations coherently modulating the polariton states. The instability regions, characterized by the presence of a coherent mechanical state, can be mapped by monitoring the polariton and the mechanical mode occupations, as shown around the $\delta E_2$ resonance in panels (b)-(f) of Fig.~\ref{Fig3}---similar features occur at $\delta E_4$. These color maps present the stationary occupations as a function of both $\delta E$ and the ratio $P/P_{\mathrm{th}}$ with $P_1=P$, $P_2=0.65P$, $P_3=0.75P$ and $P_{\mathrm{th},j}=P_\mathrm{th}$.
Panel (c)  presents the detuning between the polariton traps dressed by the optomechanical interaction as a function of the same external parameters. Several interesting features can be highlighted from these figures: i) so-called Arnold-tongues characterize the maps, separating the instability (inside) from the stability (outside) regions, the former characterized by the onset of a macroscopic occupation of the mechanical modes; ii) within these instability regions a transfer of polaritons from the ground state of the pumped trap to that of the neighbor trap is mechanically induced, even when the direct tunneling between these states is nearly zero. Note that these effects are particularly strong at the tip of the Arnold tongue, corresponding to the onset of polariton condensation. iii) A small red shift of the mechanical vibrations with increasing optical driving is evidenced by the off-centered shape of the tongues (polaritonic dressing of the mechanical states);
and iv) a locking of the polariton trap energies (main peak in the PL spectrum) at a fixed detuning that matches an integer multiple of the phonon frequency is observed within the instability regions (mechanically dressed polariton states). This locking can be connected with the universal phenomena of synchronization of non-linear dynamical systems \cite{Syncro}, also studied in the domain of quantum LF \cite{Wouters2008,Ohadi2018}.
We note, however, that in our model it is purely of optomechanical nature. In fact, non-linearities of thermal or electronic origin have not been included in the model.
\paragraph*{\textbf{Two-mode resonant cavity optomechanical system with effective quadratic coupling.}}
To bring some light to the understanding of the described physics, and to extract the more relevant solutions of what emerges as a complex dynamics, we now discuss a further simplified model.  As discussed above, the optomechanical features in the PL spectrum appear when the modes $\hat{a}_1$ and $\hat{a}_2$ are tuned to a particular energy difference (related to the phonon frequencies). This calls for a description where these two modes play the more important role. Since the excited mode is well-separated from the fundamental modes in comparison with the phonon energy, $\Delta_j=\omega_3-\omega_j\gg\Omega_n$, it is thus reasonable to assume that one can describe the dynamics with an effective reduced Hamiltonian. This can be done by means of a suitable canonical transformation.
Let us define $H'=e^{-S}He^{S}$ with $S$ an anti-hermitian operator defined by the condition
$H_\mathrm{OM}=\left[S,H_0\right]$. It can be readily seen that,
to leading order in $g_{jn}/\Delta_j$ and $\Omega_n/\Delta_j$,  and retaining only those terms involving the phonon operators,
\bea
\label{H'}
H'&=&\sum_{j=1}^3 \hbar\omega_j\,\hat{a}_j^\dagger\hat{a}_j^{}+\sum_{n} \hbar\Omega_{ n}\,\hat{b}_{n}^\dagger\hat{b}_{ n}^{ }\\
\nonumber
&&+\sum_{j=1}^2\sum_{n,m}\frac{\hbar g_{jn}g_{jm}}{\Delta}(\hat{a}_3^\dagger\hat{a}_3^{}-\hat{a}_j^\dagger\hat{a}_j^{})(\hat{b}_{n}^\dagger+\hat{b}_{n}^{})(\hat{b}_{m}^\dagger+\hat{b}_{m}^{})\\
\nonumber
&&-\sum_{n,m}\frac{\hbar g_{1n}g_{2m}}{\Delta}(\hat{a}_1^\dagger\hat{a}_2^{}+\hat{a}_2^\dagger\hat{a}_1^{})(\hat{b}_{n}^\dagger+\hat{b}_{n}^{})(\hat{b}_{m}^\dagger+\hat{b}_{m}^{})\,.
\eea
Here, for simplicity, we took $\Delta_j\pm\Omega_n\sim\Delta_j\equiv\Delta$. The second line in $H'$ reflects a coupling between the phonon displacement and the polariton mode occupations. This leads to a renormalization of the phonon energies induced by the polaritons in both the ground and excited states. When a mechanical coherent state sets in, it also leads to a modification of the polariton energies depending on the phonon occupation. The last line in $H'$, in turn,  makes explicit that there is an effective \textit{quadratic} phonon coupling between the two fundamental polariton modes of the neighboring traps, of order $G_2=g_0^2/\Delta$. As we show below this leads to a parametric instability of the phonon state, conceptually different from the self-oscillation observed in linear optomechanical systems, whenever the detuning $\delta$ equals the sum of energies of two mechanical modes of the system.

\paragraph*{\textbf{Parametric resonance.}} One can readily derive from $H'$ the equations of motion in a similar fashion as done for the full model (described in Appendix \ref{full_model}). Here we only write down the one corresponding to $\hat{x}_n$ (in the same semiclassical limit as before)
\bea
\nonumber
\Ddot{x}_n&=&-\Gamma_n\Dot{x}_n-\Omega_n^2x_n-4\Omega_n\sum_{j=1}^2\sum_{m}\frac{ g_{jn}g_{jm}}{\Delta}(a_3^*a_3^{}-a_j^*a_j^{})\,x_m\\
&&+4\Omega_n\sum_{m}\frac{ g_{1n}g_{2m}}{\Delta}(a_1^*a_2^{}+a_2^*a_1^{})\,x_m\,,
\label{xn_effective_model}
\eea
that corresponds to a set of coupled harmonic oscillators with parametric driving. The latter becomes evident when noticing that the last two terms, for $m=n$, correspond to a time dependent modulation of the phonon frequency. Moreover, the last term can become resonant at appropriate inter-trap detunings. It is important to keep in mind that this modulation provided by the polariton modes is not given but rather obtained self-consistently by solving all the equations of motion simultaneously.

To better grasp the origin of the parametric instability and derive analytical expressions for the threshold condition we look at the simpler situation where only one $p$ phonon mode exists (sufficient to describe the resonance at $\delta E_2$). In that case, Eq.~(\ref{xn_effective_model}) reduces to
\be
\Ddot{x}_n+\Gamma_n\,\Dot{x}_n+\tilde{\Omega}_n^2\,x_n=0\,,
\label{xn_single}
\ee
where
\be
\tilde{\Omega}_n^2=\Omega_n^2+4\Omega_n\frac{g_0^2}{\Delta}\left(2 n_3^{0}-\sum_{j=1}^2n_j^{0}+2\sqrt{n_1^0n_2^0}\cos{(\omega_1-\omega_2) t}\right)\,.
\label{omega_effective_zero}
\ee
Here, we took $g_1=g_2=g_0$ for simplicity, and also assumed the zero order solution (no phonon) for the polariton modes---that is $a^{0}_j(t)= \sqrt{n^{0}_j}\, e^{-\ci \omega_j t}$.
Equation (\ref{xn_effective_model}) can be cast in the form of the damped Mathieu equation~\cite{Kovacic2018}
\be
\Ddot{q} + \Gamma \Dot{q} + \left(\Omega^2 + \Omega_p^2 \cos \omega_p  t \right) q = 0\,,
\label{EQ:Mathieu0}
\ee
with $q(t)=x_n(t)$,
\bea
\nonumber
\Omega^2&=&\Omega_n^2+\frac{ 4\Omega_ng_0^2}{\Delta}\left(2 n_3^{0}-\sum_{j=1}^2n_j^{0}\right)\,,\\
\Omega_p^2&=&\frac{8\Omega_n g_0^2}{\Delta}\sqrt{n_1^0n_2^0}\,,
\label{Mathieu}
\eea
and $\omega_p=\delta=\omega_1-\omega_2$. This equation shows unstable solutions (parametric resonances) when $|\omega_p|\sim 2 \Omega/l$, with $l$ an integer number. The case $l=1$ is the standard (and most intense) double frequency parametric resonance of an oscillator, while $l>1$ describes the corresponding sub-harmonic driving conditions.  Taking  $\Omega_n$ as the fundamental confined mode at
$\Omega_0/{2\pi} = 20$~GHz this explains the experimental observation of mechanical sidebands at $\delta E_2 = -2\hbar\Omega_0$. The generalization of this result to a system with more than one phonon mode leads to resonances also at the sums of two of them (see Appendix \ref{full_model}). The resonance at $\delta E_4 = -4\hbar\Omega_0$ can thus be understood as due to a combination $\Omega_\mathrm{0} + \Omega_\mathrm{1}$. Note also that the first line of Eq.~\eqref{Mathieu} describes the variation of the mechanical frequency induced by the polaritons and explains the small shift of the center of the Arnold tongues shown in Fig. \ref{Fig3}.

A standard stability analysis of the damped Mathieu equation~\cite{Kovacic2018}
allows to find the values of the detuning where the instability sets in. Assuming that the threshold condition is met, that $\Omega\gg\Gamma$ and $\omega_1-\omega_2>0$, we get
\be
\left|\delta+2\Omega_n+\frac{ 4g_0^2}{\Delta}\left(2 n_3^{0}-\sum_{j=1}^2n_j^{0}\right)\right|\leq \frac{4g_0^2}{\Delta} \sqrt{n^{0}_1 n^{0}_2}\,.
\label{threshold_condition}
\ee
This defines the limits of the Arnold tongues shown with dotted lines in Fig.~\ref{Fig3}, which coincide very well with the numerical solutions of the full non-linear model. We stress that in this regime the phonon amplitude grows and the amplitudes of all fields have to be obtained self-consistently. Indeed, it turns out that the nonlinear back-action terms limit the amplitude of the phonon, thus avoiding the divergence of an ideal parametric oscillator. This occurs because the renormalization of the polariton energies induced by the mechanical coherent oscillation detunes the system from the exact parametric resonance. It is also worth mentioning that the same analytical condition can be derived directly from the original model described by $H$ using a perturbative approach \cite{Landau_book}.

For optimal detuning, $|\delta|= 2\Omega\sim2\Omega_n$, the threshold condition becomes
\be
\frac{4g_0^2}{\Delta \Gamma} \sqrt{n^{0}_1 n^{0}_2} > 1 \,.
\label{EQ:Threshold3}
\ee
Note that this expression retains some similarities with the threshold condition for self-oscillation in a two-mode linear optomechanical system, $C=4 \frac{N_{\mathrm{1}} |g|^2}{\kappa\Gamma_\mathrm{m}}>1$.
Firstly, it now depends on both occupations, that of the pumped and the neighbor trap. This is because what plays the role of the parametric driving in Eq.~\eqref{xn_effective_model} is the frequency beating induced by polaritons shuttling between states $1$ and $2$. Secondly, it does not depend on the cavity photon (or polariton) decay rate $\kappa$. This stems from the fact that, in the studied driven-dissipative LF, the effective $\tilde{\kappa}$ becomes zero when stimulated condensation sets in.
And thirdly, the second-order character of the optomechanical coupling,  which is mediated through virtual transitions to a delocalized excited polariton state, leads to an effective quadratic coupling of magnitude $G_2=g_0^2/\Delta$. Consequently, the threshold for a parametric resonance instability is changed with respect to the linear expression by a factor of order $\Delta/\kappa$. If we take for $\kappa$ the inverse of the cavity photon lifetime ($\sim 10$~ps), this number is rather small, of the order of $\sim 6$. The implication is that the described mechanical instability threshold power is similar to that for an equivalent linear system, despite the fact that the direct linear coupling between isolated traps is hugely decreased due to the very small overlap integrals (the threshold power for self-oscillations in the linear optomechanical regime in our device is estimated to be several orders of magnitude larger due to this effect).

As mentioned before, we have neglected the $s$-phonon modes in our analysis. This is essentially correct for the sake of understanding the origin of the mechanical instability as it turns out that only the $p$-phonon modes are responsible for the emergence of a coherent mechanical state. Nevertheless, it is important to point out that we have observed that, once the $p$-modes becomes unstable, they trigger the oscillation of the $s$-modes through the modulation of the population of the polaritonic modes at the proper frequency. When this occurs all phonon modes can become macroscopically populated and lead to enhanced sidebands.
\section{Discussion and outlook}
The phenomena we have reported and termed as optomechanical parametric oscillation, is different from other realisations of quadratic Hamiltonians in cavity optomechanics in several relevant aspects, namely: i) it applies to optomechanical crystals that are driven, or populated, by an exciton-polariton light fluid, ii) concomitant with that, no resonant driving is performed: the self-oscillations arise from the intrinsic dynamics of a non-resonantly excited driven-dissipative system; iii) the optomechanical coupling is essentially purely quadratic, with linear terms negligible for all practical purposes; iv) it is a fully resonant two-mode light-fluid system, in which these two involved modes self-tune to the parametric resonance at double frequencies; and v) because of the super-high mechanical frequencies involved ($20$~GHz~$\sim 1$~K and $60$~GHz~$\sim 3$~K), the mechanical system is very close to the quantum ground state at the working standard cryogenic temperatures ($\sim 4$~K). These peculiarities lead to some consequences that we discuss next.
\paragraph*{\textbf{Self-oscillation vs. parametric resonance.}}
A parametric resonance as described here is similar but conceptually different from the phenomena of self-oscillation~\cite{Jenkins2013} present in cavity systems with linear optomechanical coupling \cite{RMP}. The phenomena of self-oscillation is a property of  dynamical systems characterized by a driving force that is controlled by the oscillation, acting in phase with the velocity. These phenomena can be linked also to retarded restoring forces. No explicit time dependence of the force is, however, required. In cavity optomechanical systems it leads to an additional contribution determined by the optical force to the term in the oscillator equation that is proportional to $\Dot{q}$. Above a certain threshold this causes a negative damping, leading to an amplitude of the oscillation that grows exponentially with time, being ultimately limited by other non-linear effects. Parametric resonance resembles self-oscillation in that the growth of the amplitude is also exponential in time but, in this case, the equation of motion has an explicit time-dependence \cite{Jenkins2013}. In fact, it requires a perturbation affecting the oscillator frequency with a harmonic time dependence tuned to $|\omega_p|\sim 2 \Omega/l$. In our case, this harmonic driving is intrinsically generated by the beating of a light-fluid between two detuned modes belonging to neighboring traps. In this way an optomechanical parametric oscillator (OMPO) is realised, which mimics quite precisely its equivalent in optics, the optical parametric oscillator (OPO).
\paragraph*{\textbf{Limit cycles and chaos.}}
The experimental observation of symmetrical mechanical sidebands is a manifestation of the emergence of a periodic limit cycle with the parametric oscillation characteristics just discussed. The numerical modelling of the system also evidences a more complex dynamics when the multi-variable parameter space is investigated in a broader range of attainable conditions. Different phases emerge within the optomechanical instability (Arnold tongues). Notably, besides the periodic limit cycles stable within  part of the parameter space, we have observed  at larger excitation powers  and within the Arnold tongues that there are regions  characterised by chaotic-like regimes. Such chaotic phases have indeed been identified in other optomechanical systems \cite{Navarro2017,Wu2017,Figueiredo2020}, including resonantly driven systems with quadratic coupling \cite{Zhang2017}.  The numerical simulations show that while the polariton decay rate $\kappa$ does not affect the threshold for parametric oscillation (as discussed above), it does determine the peculiar geography of the different phases within the Arnold tongues.
\paragraph*{\textbf{Optomechanically induced tunneling.}}
The reported optomechanically induced tunneling mechanism, mediated by a non-resonant excited extended state, is different from other relevant tunneling processes identified in trapped interacting bosonic condensates. The latter typically require a direct coupling between the involved initial and final states \cite{Smerzi1997,Albiez2005,Zollner2008,Lode2012,Lagoudakis2010,Abbarchi2013}. Based on the described optomechanically induced tunneling, phonons could be used for the control of operations in quantum simulators based on light fluids \cite{Hartmann2006,Kalinin2020,Alyatkin2020,Gosh2020}.
The optomechanical coupling, largely unexplored in the domain of exciton-polariton light fluids, can be tailored to display a rich variety of physical phenomena that could be relevant for quantum technologies.
\paragraph*{\textbf{Squeezing and two-phonon coherent states}}
The physics we described here is conceptually quite similar to optical parametric oscillation in photonics by which photons are generated at frequency $\Omega_0$ through non-linear processes induced by driving at $2\Omega_0$ in appropriate crystals lacking  inversion symmetry.
Indeed, the Hamiltonian for such systems  takes the form $H_\mathrm{OPO}=\lambda(\hat{b}^{2}\hat{a}^\dagger+\hat{b}^{\dagger 2}\hat{a}^{})$, with operators $a$ ($b$) standing for photons of frequency $2\Omega_0$ ($\Omega_0$).
The photonic driving field $a$ is typically assumed large and thus taken as a complex number $a\rightarrow i\frac{r}{2} e^{-\ci 2\Omega_0t}$. With this, the time evolution operator of a driven OPO system, in the appropriate reference frame, essentially  becomes the squeezing unitary operator $S(z)=\exp\left[\frac{ z}{2}\left(\hat{b}^2-\hat{b}^{\dagger2}\right)\right]$, where $z=\lambda r/\hbar$ here is taken as a real number.
A similar procedure can be followed for our polariton system with effective quadratic optomechanical interactions when operating at the resonance $\delta E_2 \sim -2\hbar \Omega_0$, corresponding to the coupling with pairs of identical $\hbar \Omega_0$ phonons.
To make a full parallel with the OPO Hamiltonian, one simply has to identify the parameters $\frac{r}{2}=\sqrt{n^{0}_1 n^{0}_2}$ and $\lambda=\frac{\hbar g^2}{\Delta} $ in Eq. \eqref{H'}.
Under this assumptions, the mechanical wave function resulting from this polariton driving on the phonon vacuum $\left|0\right>$ (the $20$~GHz mode is very close to its quantum ground state at our working temperature of 4~K, with an initial thermal occupation around 4) can be expressed as a squeezed vacuum, a very energetic state resulting from the coherent superposition of even number states, $S(z) \left|0\right>$ is identified as a two-phonon coherent state. In this sense, our system behaves as a fully resonant optomechanical parametric oscillator.
In the actual device the coherent phonon population is limited both by non-linearities of the full Hamiltonian describing the system, and by the residual dephasing of the polariton condensates. The coherence time of the polariton condensates has been experimentally determined to be around $600$~ps. Preliminary numerical simulations including this decoherence show that it does not modify substantially the dynamics of the system, which thus seems to be mostly limited by the intrinsic non-linearities of the light-fluid optomechanics.
We thus envision that properly designed systems might take advantage of this effective quadratic optomechanical coupling to produce entangled phonon pairs and well defined squeezed phonon states on demand.
\begin{acknowledgments}
We acknowledge partial financial support from the ANPCyT-FONCyT (Argentina) under grants PICT-2015-1063, PICT-2018-03255, PICT-2016-0791 and PICT 2018-1509, CONICET grant PIP 11220150100506, SeCyT-UNCuyo grant 06/C603, from the  German Research Foundation (DFG) under grant 359162958, and the joint Bilateral Cooperation Program between the German Research Foundation (DFG) and the Argentinian Ministry of Science and Technology (MINCyT) and CONICET. AF thanks the Alexander von Humboldt Foundation and the Paul Drude Institut for the support and hospitality while this work was completed.
\end{acknowledgments}

\appendix
\section{Device fabrication\label{Fab}}
The studied device consists of an microstructured polaritonic microcavity with arrays of $\mu$m-sized intracavity traps able to confine polaritons in three dimensions. The microcavity is created by patterning (Al, Ga)As in-between growth steps by molecular beam epitaxy (MBE). The fabrication process is as follows: First a $4.43\,\mu$m thick lower distributed Bragg reflector (DBR) consisting of $36$ $\lambda/4$ ($\lambda$ is the optical wavelength) pairs of Al$_{0.15}$Ga$_{0.85}$As/Al$_{x}$Ga$_{1-x}$As was grown on $350\,\mu$m GaAs substrate. The Al composition $x$ of the lower DBR is continuously reduced from $0.80$ in the first stack to $0.45$ in the last stack. Then, the first $120$ nm of the Al$_{0.30}$Ga$_{0.70}$As microcavity spacer were deposited including six $15$-nm-thick GaAs quantum wells (QWs) placed at the antinode positions of the microcavity optical mode. The structure was subsequently capped by a $170$-nm-wide Al$_{0.15}$Ga$_{0.85}$As layer spacer. This last layer protects the QWs for the next step, when the unfinished-sample is removed from the MBE chamber and patterned by means of photolithography and wet chemical etching. Here, mesas with a nominal height of $12$ nm of different shapes in the exposed spacer layer were created inducing a lateral modulation of the cavity thickness and, therefore, of the cavity energy in the final structure. Followed this, the sample was reinserted into the MBE system, cleaned by exposure to atomic hydrogen, and overgrown with a $\lambda/4$ Al$_{0.15}$Ga$_{0.85}$As layer. Finally, the upper DBR was grown by 20 $\lambda/4$ pairs of Al$_{0.15}$Ga$_{0.85}$As/Al$_{0.75}$Ga$_{0.25}$As.

The etched regions result in a blueshift of the optical cavity mode in the etched areas by $9$ meV ($4.5$ nm) with respect to the non-etched regions. The upper surface of the etched layer spacer corresponds to a node of the optical cavity mode of the whole structure. In this way, potential impact of roughness or impurities introduced by the ex situ patterning on optical properties of the structure was minimized. Furthermore, the shallow patterned layer is located more than $140$ nm above the QWs, so that they remain unaffected by the processing.

The sample was designed to be in the strong coupling regime at low-temperature ($\sim5$K) both in the etched and nonetched regions, leading to microcavity polaritons in these two regions with different energies and photon/exciton contents. The lateral modulation was used to create 3D (square/dots potential) confinement in non-etched areas surrounded by etched barriers. A detailed characterization of the reflection and photoluminescence and the characterization of the potential profile of the polaritonic traps in this sample is presented in Ref. \cite{Kuznetsov2018}.

\section{Polariton modes: the Gross-Pitaevskii equation \label{GP}}
The effective confinement potential for the polaritons, and the corresponding modes shown in Figs.~\ref{Fig1}(c) and \ref{Fig1}(d) were obtained using an effective Gross-Pitaeskii equation that takes into account both the blue shift induced by the repulsive interactions with the exciton reservoir and the saturation of the Rabi splitting \cite{Mangussi2021}. Namely,
\begin{equation}
\label{Effective_LP_Equation}
 i \hbar \dfrac{\partial\psi}{\partial t} = \left[-\frac{\hbar^{2}}{2 m_{LP}} \nabla^{2} + V_{LP}(\bm{r})+i\left(\frac{R\, P(\bm{r})}{\gamma_R+R |\psi|^2}-\kappa\right)\right] \psi\,,
\end{equation}
where $\psi(\bm{r},t)$ is the complex field describing the lower polaritons and we have used the adiabatic approximation for the exciton reservoir \cite{CarusottoRMP2013}.  Equation \eqref{Effective_LP_Equation} is a very good approximation when describing the confined polaritonic levels $s$ and $p$. In addition, in our particular case, and for the purposes of describing only the energy of the polaritonic modes (not their occupation) and the effective potential, Eq. \eqref{Effective_LP_Equation} can be further simplified by ignoring the last term \cite{Mangussi2021}.
The effective potential is defined as
\begin{equation}
\label{Effective_potential}
V_{LP}(\bm{r},n_{R})=\frac{1}{2}\left[ E_{C}+E_{X}-\sqrt{\Omega^{2}+\Delta^{2}}\right]\,,
\end{equation}
with $E_{C}(\bm{r})=\Delta_{0}+V_{C}(\bm{r})$ the photonic potential of the trap, $\Delta_{0}$ the bare detuning, $E_{X}(\bm{r},n_{R})=g_{X}n_{R}(\bm{r})$ the exciton energy ($g_X\approx 6\mu$eV$\mu$m$^2$) and  $\Delta(\bm{r},n_{R})=E_{C}(\bm{r})-E_{X}(\bm{r},n_{R})$ the effective detuning. Here we have ignored the contribution to $E_X$ from the repulsive interaction among the lower polaritons. This is fine in this case as the parameters are such that they have a large photonic component.

The saturation of the Rabi splitting with increasing population of the reservoir is described  as
\begin{equation}
\label{Rabi}
\Omega(\bm{r},n_{R})=\frac{\Omega_{0}}{\sqrt{1+\frac{n_{R}(\bm{r})}{n_{\mathrm{Sat}}}}},
\end{equation}
where $\Omega_{0}$ is the Rabi splitting at zero carrier density and $n_{\mathrm{Sat}}\approx 3\times 10^3\mu$m$^{-2}$.
The effective mass is approximated as
\begin{equation}
\label{m_LP}
\frac{1}{m_{LP}}=\frac{|X(\bm{r}_0,n_{R})|^2}{m_{X}}+\frac{|C(\bm{r}_0,n_{R})|^2}{m_{C}}\,,
\end{equation}
where $\bm{r}_0$  corresponds to the center of the pumped trap and the spatially dependent  Hopfield coefficients are
\bea
\nonumber
|X(\bm{r},n_{R})|^2 &=&\frac{1}{2}\left(1+\frac{\Delta(\bm{r},n_{R})}{\sqrt{\Omega(\bm{r},n_{R})^{2}+\Delta(\bm{r},n_{R})^{2}}}\right)\,, \\
|C(\bm{r},n_{R})|^2 &=& 1-|X(\bm{r},n_{R})|^2 \,.
\label{Hopfield}
\eea
Note that we have explicitly taken into account the dependence of the parameters with the density of carrier in the reservoir.
The density of excitons in the reservoir as a function of the external pump power is estimated as
\begin{equation}
\label{n_R}
n_{R}(\bm{r}) = \frac{P(\bm{r}) \tau_{R} \alpha}{\hbar \omega_{L} \, 2 N_{QW}}\,,
\end{equation}
where $P(\bm{r})$ is the pump power per unit area, $\tau_{R}$ is the effective lifetime of the exciton in the reservoir, $\alpha$ is the total effective absorption coefficient of the QWs, $\hbar \omega_{L}$ is the energy of the non resonant pumping laser, $N_{QW}$ is the number of quantum wells and the factor $2$ accounts for the dominant role of the triplet interactions.
We assume a Gaussian shape for the pump given by
\begin{equation}
P(\bm{r})=\frac{P_{0}}{2 \pi \sigma_{p}^{2}} \exp\left[-\frac{(\bm{r}-\bm{r}_{p})^{2}}{2\sigma_{p}^{2}}\right]
\end{equation}
In the simulations we use an effective value for the standard deviation $\sigma_{p} \approx 3 \mu$m, and change the position of the spot ($\bm{r}_{p}$) to reproduce the particular experimental situation. %
The values for $\Omega_{0}$, $\Delta_{0}$ and the cavity parameters $m_{C}$ and $V_{C}(\bm{r})$ were obtained by fitting. The photonic cavity potential $V_{C}(\bm{r})$ was simulated following Refs. \cite{Kuznetsov2018,Mangussi2021}. The optimal results were obtained using $\Omega_{0}=6.0$meV and $E_{C}=\Delta_{0}=-10.5 \, \text{meV}$ and $E_{C}=\Delta_{0}+U=5.5 \, \text{meV}$ for the non etched and etched regions, respectively. Here $U=16$ meV is the potential barrier for photons, generated by the difference in the thickness of the cavity spacer between the two regions.

 \section{Optomechanical coupling constant \label{gfactor}}

To estimate the linear on-site optomechanical coupling factor $g_0$, corresponding to processes that couple polariton levels within the same trap, we begin by considering the effective exciton-mediated optomechanical coupling reported in Ref.~\cite{Kuznetsov2021} for similar individual polariton traps as those investigated in this work. This coupling results mainly from a deformation-potential interaction modulated by intense electrically generated surface acoustic waves (SAWs). A value of $g^\mathrm{eff}_\mathrm{om}/2\pi \sim 50$~THz/nm was obtained, which accounts for the change of polariton energy per unit of acoustic displacement \cite{Kuznetsov2021}. This later parameter is related to the actual on-site optomechanical coupling constant by $g^\mathrm{eff}_0=g^\mathrm{eff}_\mathrm{om}\,x_\mathrm{zpf}$\,\cite{RMP}, where $x_\mathrm{zpf}$ corresponds to the displacement induced by the zero point fluctuations. For a similar structure of $\sim2\,\mu$m lateral size, this value has been estimated to be roughly $x_\mathrm{zpf}\sim 0.5$\,fm~\cite{Villafane2018}. Consequently, $g^\mathrm{eff}_{0}/2\pi \sim 50$~MHz for this system, which represents a  very large value compared to other reported optomechanical systems that only account for an optical radiation back-action mechanism based on radiation pressure interaction\,\cite{RMP}. The Hopfield coefficient for this cavity polariton trap system was of the order of $|X|^2\sim 0.7$, and the reported structure had only one embedded QW~\cite{Kuznetsov2021}.

The structure investigated in this work, has six cavity embedded QWs instead of one, proportionally increasing the corresponding interaction of the involved fields. For high excitation powers used for exciting the traps analyzed here, the excitonic Hopfield coefficient is estimated to be around $|X|^2\sim 0.05$ \cite{Mangussi2021}, i.e. the involved polariton states have a large photonic component. Considering these two differences, the $g_0$ would be roughly a factor of $6$ times larger and a factor $0.05/0.7$ smaller with respect to the above obtained $g^\mathrm{eff}_{0}$. Therefore, the on-site optomechanical coupling factor for the present work, resulting from a deformation-potential interaction, can be estimated to be
$g_{0}/2\pi\sim 20\,\mathrm{MHz}$.
\section{Simplified polariton-phonon model\label{full_model}}
We present here the effective model used to describe the polaritons optomechanical coupling, but now explicitly including the $s$-phonon modes that were ignored in the main text for the sake of simplicity. The full Hamiltonian then reads
\bea
\nonumber
H&=&\sum_{j=1}^3 \hbar\omega_j\,\hat{a}_j^\dagger\hat{a}_j^{}+\sum_{n} \hbar\Omega_{ n}\,\left(\sum_{j=1}^2 \hat{d}_{j n}^\dagger\hat{d}_{j n}^{}+\hat{b}_{n}^\dagger\hat{b}_{ n}^{ }\right)\\
&&+\sum_{j=1}^2\sum_n \hbar g_{jn}(\hat{a}_j^\dagger \hat{a}_3^{}+\hat{a}_3^\dagger \hat{a}_j^{})(\hat{b}_{n}^\dagger+\hat{b}_{n}^{})\\
\nonumber
&&+\sum_n\sum_{j=1}^2 (\hbar \bar{g}_n\, \hat{a}_j^\dagger\hat{a}_j^{}+\hbar \bar{g}_{jn}\,\hat{a}_3^\dagger\hat{a}_3^{} ) (\hat{d}_{j n}^\dagger+\hat{d}_{j n}^{})\,,
\eea
where: i)   $\hat{a}_j^{\dagger}$ ($\hat{a}_j$) creates (annihilates) a polariton in the $j$-mode with energy $\hbar\omega_j$ where  $j=3$ refers to the excited mode; ii) $\hat{d}_{j n}^{\dagger}$ and $\hat{b}_{n}^{\dagger}$  ($\hat{d}_{j n}$ and $\hat{b}_{n}$) create (annihilate) a phonon in the ${n}$-mode with $s$-like and $p$-like symmetry, respectively, and energy $\hbar\Omega_{n}$---here we ignored the very small energy difference between the $s$ and $p$ modes. The index $n$ labels the fundamental and the overtone mechanical modes so that, for example,  $\Omega_1=3\Omega_0=60$GHz. We assume the $s$ phonon modes to be localized on each cavity (hence the additional index $j$) and  the $p$ modes to be extended and so shared between cavities.
Note that $H_\mathrm{OM}$ does not include a direct coupling between $\hat{a}_1$ and $\hat{a}_2$ but only with the excited mode.

From the above Hamiltonian it is straightforward to derive the equations of motion for $\hat{a}_j$ and for the dimensionless phonon position operators, $\hat{x}_n=\hat{b}_{n}^\dagger+\hat{b}_{n}^{}$ and $\hat{y}_{jn}=\hat{d}_{jn}^\dagger+\hat{d}_{jn}^{}$. In the semiclassical approximation where the bosonic operators are replaced by complex functions we obtain the following set of equations
\bea
\nonumber
i\Dot{a}_j
&=&\left(\omega_j+\sum_n \bar{g}_n y_{jn}\right)a_j+\sum_n g_{j
n}\, x_n a_3+F_j(a_j)\,,\,\,\, j\neq3\\
\nonumber
i\Dot{a}_3 &=&\left(\omega_3+\sum_{j=1}^2\sum_n \bar{g}_{jn}y_{jn}\right)a_3+\sum_{j=1}^2\sum_n g_{jn}\,x_n a_j+F_3(a_3)\,,\\
\nonumber
\Ddot{x}_n&=&-\Omega_n^2x_n-\Gamma_n\Dot{x}_n-2\Omega_n\sum_{j=1}^2 g_{jn}(a_ja_3^*+a_3a_j^*)\,,\\
\Ddot{y}_{jn}&=&-\Omega_n^2y_{jn}-\bar{\Gamma}_n\Dot{y}_{jn}-2\Omega_n\,\left(\bar{g}_{n}|a_j|^2+\bar{g}_{jn}|a_3|^2\right)\,.
\label{Eq:full_model}
\eea
Here we added a phenomenological term $F_j(a_j)$ to account for the incoherent driving of the polariton modes induced by the excitons reservoir. Based on the  GP equation, and for the purpose of introducing a stationary population of the polaritonic modes,  we use the following simplified expression for it $F_j(a_j)=i(\bar{R}P_j/(\gamma_R+\bar{R}|a_j|^2)-\kappa)a_j/2$, which immediately leads to the effective decay rate $\tilde{\kappa}_j$ introduced in the main text.
This allow us to describe the condensation of each mode, when the incident power from the reservoir to mode $j$, $P_j$, is larger than the threshold power $P_\mathrm{th}=\kappa \gamma_R/\bar{R}$. In absence of the optomechanical effects each mode would condense to an occupation $n^{0}_j = (P_{j}/P_\mathrm{th}-1)n_0$ with $n_0=\gamma_R/\bar{R}$.
Additionally, we also included a dissipative term proportional to $\Gamma_{n}$ and $\bar{\Gamma}_n$ (frequency linewidth) in the equations for $x_n$ and $y_{jn}$, respectively, to account for the decay of the phonon modes.

The full model can be solved numerically. As mentioned in the main text, while the instability is caused by the $p$ phonon that couples $a_1$ and $a_2$ through the excited states, once this occurs the $s$ phonons are also excited. This leads to a renormalization of the Arnold tongues and a rich interplay between phonons, whose analysis is beyond the scope of the present work. 

The description presented above, though conceptually simple and relatively easy to solve numerically, it is still too complex as to easily grasp the more relevant solutions. Since the excited polariton mode is well separated from the fundamental modes in comparison with the phonon energy, $\Delta_j=\omega_3-\omega_j\gg\Omega_n$, one can describe the dynamics with an effective reduced Hamiltonian. This can be done by means of a suitable canonical transformation.
Defining $H'=e^{-S}He^{S}$ with $S$ given by
\be
S=\sum_{j=1}^2\sum_n \hat{a}^\dagger_j \hat{a}_3 \left(\frac{g_{jn}}{\omega_3-\omega_j+\Omega_n} \hat{b}_n+ \frac{g_{jn}}{\omega_3-\omega_j-\Omega_n}\hat{b}_n^\dagger\right)-h.c.\,,
\ee
one gets, to leading order in $g_{jn}/\Delta_j$ and $\Omega_n/\Delta_j$  and retaining only those terms involving the phonon operator,
\bea
\nonumber
&&\\
\nonumber
H'&=&\sum_{j=1}^3 \hbar\omega_j\,\hat{a}_j^\dagger\hat{a}_j^{}+\sum_{n} \hbar\Omega_{ n}\,\left(\sum_{j=1}^2 \hat{d}_{j n}^\dagger\hat{d}_{j n}^{}+\hat{b}_{n}^\dagger\hat{b}_{ n}^{ }\right)\\
\nonumber
&&+\sum_{j=1}^2\sum_n (\hbar \bar{g}_n\, \hat{a}_j^\dagger\hat{a}_j^{}+\hbar \bar{g}_{jn}\,\hat{a}_3^\dagger\hat{a}_3^{} ) (\hat{d}_{j n}^\dagger+\hat{d}_{j n}^{})\\
\nonumber
&&+\sum_{j=1}^2\sum_{n,m}\frac{\hbar g_{jn}g_{jm}}{\Delta}(\hat{a}_3^\dagger\hat{a}_3^{}-\hat{a}_j^\dagger\hat{a}_j^{})(\hat{b}_{n}^\dagger+\hat{b}_{n}^{})(\hat{b}_{m}^\dagger+\hat{b}_{m}^{})\\
\nonumber
&&-\sum_{n,m}\frac{\hbar g_{1n}g_{2m}}{\Delta}(\hat{a}_1^\dagger\hat{a}_2^{}+\hat{a}_2^\dagger\hat{a}_1^{})(\hat{b}_{n}^\dagger+\hat{b}_{n}^{})(\hat{b}_{m}^\dagger+\hat{b}_{m}^{})\,,\\
\eea
where we used that $\Delta_j\pm\Omega_n\sim\Delta_j\equiv\Delta$.
Eq. \eqref{xn_effective_model} in the main text is derived from $H'$, which can be cast in the form of a Mathiew equation (cf. Eq. \eqref{EQ:Mathieu0}). This equation shows unstable solution  when $|\omega_p|\sim 2 \Omega$.
The stability analysis is rather standard, see for instance Refs. [\onlinecite{Landau_book,Kovacic2018}]: the condition for finding an unstable solution is given by
\be
\Omega_p^2 > \sqrt{\Gamma^2 \omega_p^2  + 4\left(\Omega^2-\frac{\omega_p^2}{4}\right)^2}\,,
\label{EQ:Threshold}
\ee
from where Eqs. \eqref{threshold_condition} and \eqref{EQ:Threshold3} can be obtained.

So far we considered a single $\Omega_0$ phonon mode, which is enough to account for the resonance observed at $\delta E_2$---for that resonance, adding the $\Omega_1$ phonon leads to a complex interplay when the $g_n$ couplings are similar. On the contrary, to describe the effects observed at $\delta E_4$ one must consider the $\Omega_1$ phonon.
That is, one has to fully consider the two coupled oscillators that are associated with two $p$ phonons as describe in Eq.~(\ref{xn_effective_model}). As before, taking the zero-order solution for the polariton modes one can map the equations to the ones of two damped oscillators driven by a crossed parametric excitation of frequency $\omega_p$,
\bea
 \Ddot{q}_1  + \Gamma_1  \Dot{q}_1  + \Omega^2 q_1 + \Omega_p^2\cos(\omega_p  t) q_3 &=& 0 \,,\nonumber \\
\Ddot{q}_3  + \Gamma_3  \Dot{q}_3 + 9 \Omega^2 q_1 + 3\Omega_p^2\cos (\omega_p  t) q_1 &=& 0 \,,
\label{EQ:Coupled2}
\eea
where the coordinates $q_1$ and $q_3$ describe the oscillator having natural frequency $\Omega$ and $3\Omega$, respectively, $\omega_p=\omega_1-\omega_2$ and  the amplitude of the drive is given by  $\Omega_p^2= 8 \Omega \frac{g_{1n} g_{2m}}{\Delta} \sqrt{n^{0}_1 n^{0}_2}$, with $n$ and $m$ labeling the $p$-phonon modes with frequency $20$~GHz and $60$~GHz, respectively. The frequency $\Omega$ can be assumed shifted as in the $\delta E_2$ case.  By applying the two variable expansion method \cite{Kovacic2018} for  $\omega_p = 4\Omega$, i.e. the frequency sum of the two coupled oscillators, we obtain the following condition for phonon instability at optimal detuning
\be
4 \frac{g_{1n} g_{2m}}{\Delta \sqrt{\Gamma_{1} \Gamma_{2}}} \sqrt{n^{0}_1 n^{0}_2} > 1
\label{EQ:Threshold3b}
\ee
This is essentially the same condition found for a single oscillator. We also obtain an expression as a function of the detuning near $\omega_p=4\Omega$,
\be
\Omega_p^2 > \sqrt{\frac{\Gamma_1^2 \omega_p^2}{4}  + 16\left(\Omega^2-\frac{\omega_p^2}{16}\right)^2}
\label{EQ:ThresholdB}
\ee
where we have assumed for simplicity that $\Gamma_1=\Gamma_3$.


\end{document}